\documentclass[journal,twoside,web]{IEEEtran}
\usepackage{generic}
\usepackage{cite}
\usepackage{hyperref}
\usepackage{amsmath,amssymb,amsfonts}
\usepackage{algorithmic}
\usepackage{graphicx}
\usepackage{textcomp}
\usepackage{caption}
\usepackage{subcaption}
\usepackage{bm}
\usepackage[utf8]{inputenc}
\usepackage[english]{babel}
\usepackage{lscape}
\usepackage[table]{xcolor}
\usepackage{multirow}
\usepackage{multicol}
\usepackage{array}
\usepackage{tabularx}

\newcolumntype{Y}[1]{>{\raggedright\arraybackslash}m{#1}}

\usepackage{booktabs}
\usepackage{threeparttable}

\begin{document}
\title{Neuromorphic Neuromodulation: \\{Towards the next generation of on-device AI-revolution in electroceuticals}}
\author{Luis Fernando Herbozo Contreras, Nhan Duy Truong \IEEEmembership{Member, IEEE}, Jason K. Eshraghian \IEEEmembership{Member, IEEE}, Zhangyu Xu, Zhaojing Huang, Armin Nikpour, Omid Kavehei \IEEEmembership{Senior Member, IEEE}
\thanks{This work has been submitted to the IEEE for possible publication. Copyright may be transferred without notice, after which this version may no longer be accessible.} 
\thanks{Luis Fernando Herbozo Contreras, Nhan Duy Truong, Zhangyu Xu, Zhaojing Huang and Omid Kavehei are with the School of Biomedical Engineering, The University of Sydney, NSW 2006, Australia (e-mail: omid.kavehei@sydney.edu.au)} 
\thanks{Nhan Duy Truong and Omid Kavehei are with BrainConnect Pty Ltd, Darlington, NSW 2008, Australia.}
\thanks{Jason K. Eshraghian is with the Department of Electrical and Computer Engineering, University of California, Santa Cruz 95064, USA}
\thanks{Armin Nikpour is with Central Clinical School, The University of Sydney, NSW 2006, Australia}} 

\maketitle

\begin{abstract}
Neuromodulation techniques have emerged as promising approaches for treating a wide range of neurological disorders, precisely delivering electrical stimulation to modulate abnormal neuronal activity. While leveraging the unique capabilities of artificial intelligence (AI) holds immense potential for responsive neurostimulation, it appears as an extremely challenging proposition where real-time (low-latency) processing, low power consumption, and heat constraints are limiting factors. The use of sophisticated AI-driven models for personalized neurostimulation  depends on back-telemetry of data to external systems (e.g. cloud-based medical mesosystems and ecosystems). While this can be a solution, integrating continuous learning within implantable neuromodulation devices for several applications, such as seizure prediction in epilepsy, is an open question. We believe neuromorphic architectures hold an outstanding potential to open new avenues for sophisticated on-chip analysis of neural signals and AI-driven personalized treatments. With more than three orders of magnitude reduction in the total data required for data processing and feature extraction, the high power- and memory-efficiency of neuromorphic computing to hardware-firmware co-design can be considered as \textit{the} solution-in-the-making to resource-constraint implantable neuromodulation systems. This perspective introduces the concept of \textit{Neuromorphic Neuromodulation}, a new breed of closed-loop responsive feedback system. It highlights its potential to revolutionize implantable brain-machine microsystems for patient-specific treatment.

\end{abstract}


\section{Introduction}
\IEEEPARstart{E}{lectrical} brain stimulation has evolved significantly over the past more than half a century. It started in the 50s when it was found that emotional responses can be triggered by electrical brain stimulation~\cite{delgado1954stim}. Since then, there has been an increasing number of studies on the safety of brain stimulation ~\cite{Goddard1969stim, gordon1990parameters} and its applications as therapy of intractable epilepsy~\cite{osorio2001neurostim, litt2003translating, piper2022towards}, spinal cord injury~\cite{Wenger2014clnm}, psychiatric illness~\cite{Lo2017clnm}, Parkinson's disease~\cite{Kim2020clnm}, dystonia~\cite{kupsch2006dystonia}, refractory depression~\cite{mayberg2005depression} and Alzheimer's disease~\cite{laxton2010alzheimer, lozano2016alzheimer}. However, there is yet to be an effective, scalable, personalized, and truly responsive stimulation solution for refractory epilepsy or neurological diseases in general.

\begin{figure*}
\centering
\includegraphics[width=0.80\textwidth]{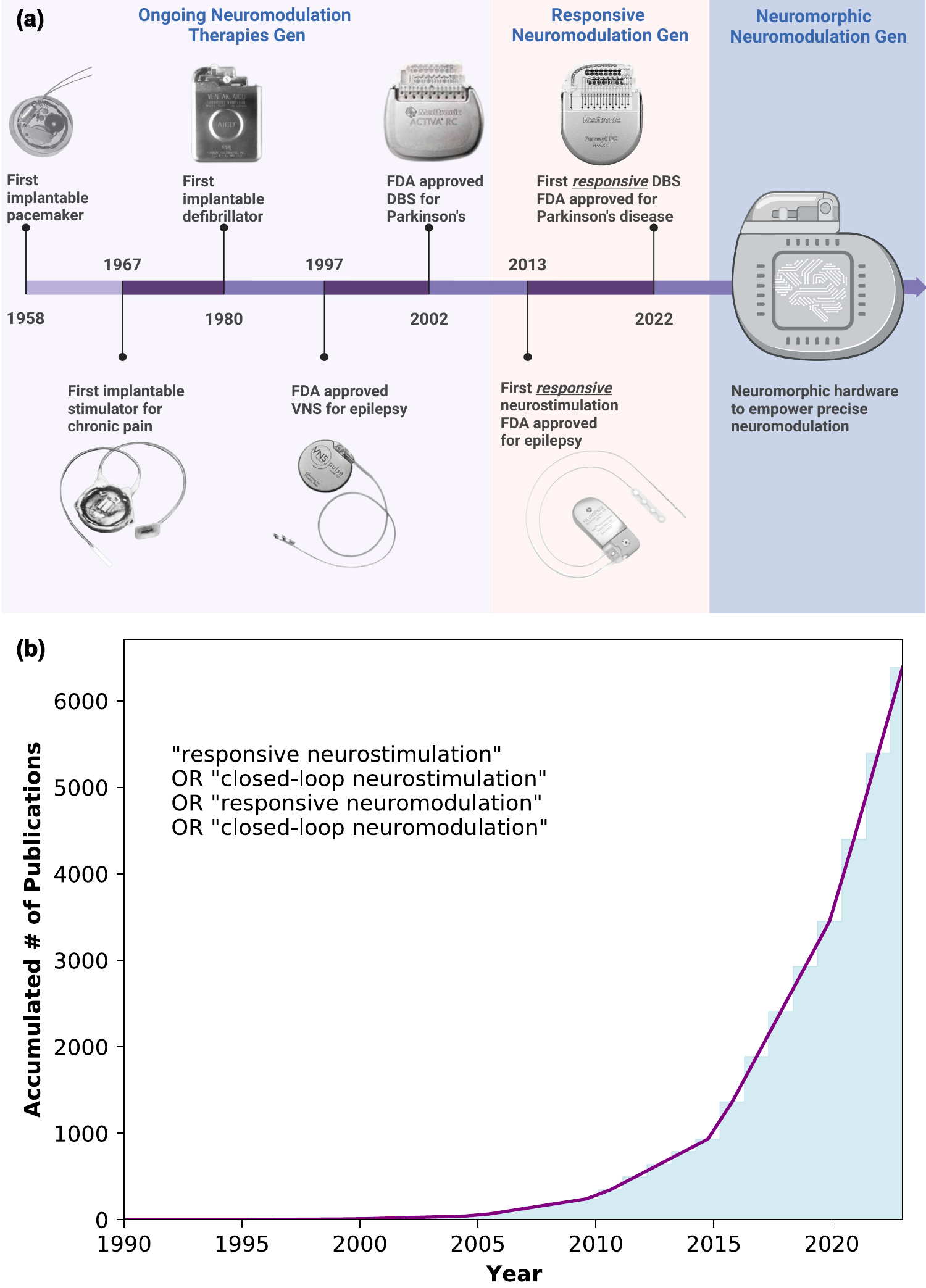}
\caption{{(a) History of neurostimulation devices and future directions ~\cite{Kerzenmacher2013firstpace, shealy1970dorsal, Mirowski1980defibrillator, Lulic2009vns, Medtronic2021dbs, suarez2017earlier, sun2014rns, jimenez2021device}}. The first generation of neuromodulation devices primarily involved the delivery of constant electrical stimulation to targeted brain regions. Responsive stimulation generation represents a significant leap forward, incorporating closed-loop systems that dynamically adjust stimulation parameters based on real-time feedback from neural activity or physiological markers. Inspired by brain computing, the future generation should be focused on neuromorphic neuromodulation, which holds great potential for revolutionary and precise therapeutic interventions.
(b) Accumulated number of publications in responsive or closed-loop neurostimulation. An evident gap exists in the field of closed-loop systems about the requirement for on-chip devices capable of continuous learning.}
\label{neuromorphic:history}
\end{figure*}

The market share of neurostimulation devices was more than US\$6B in 2020 and is projected to pass US\$11B by 2026~\cite{MarketWatch2021}. Key manufacturers of neurostimulation devices include Medtronic, Boston Scientific, Abbott, LivaNova, Nevro, NeuroPace, Beijing Pins, and Synapse Biomedical. Fig.~\ref{neuromorphic:history} depicts a glimpse of the history of implantable neurostimulation devices and the trend in advanced neurostimulation. Although it does not perform neurostimulation, we consider the first pacemaker~\cite{Kerzenmacher2013firstpace} the first important milestone on the roadmap, as it shares the same core idea: electrical stimulation. A decade after the first pacemaker, in 1967, the first implantable stimulation device was introduced for chronic pain relief. Since then, neurostimulation has shown consistent effectiveness in reducing chronic pain~\cite{Hofmeister2020pain}. This is followed by the first implantable defibrillator reported in 1980~\cite{Mirowski1980defibrillator}.

Neurostimulation has been explored for its potential as a treatment or therapy for other diseases such as epilepsy, Parkinson's disease, Alzheimer's disease, and spinal cord injury. The year 1997 marks the first FDA-approved vagus nerve stimulation (VNS) device in treating intractable epilepsy~\cite{Lulic2009vns}. The device, NeuroCybernetic Prosthesis, is based on the finding that stimulating the vagus nerve modulates cortical activity via thalamocortical pathways, though the precise mechanism is not yet fully understood~\cite{rao2021chronic}. Deep brain stimulation (DBS) was first used in 1980 for the reduction of tremors~\cite{Brice1980dbs} and has since become an effective treatment of Parkinson's disease with impressive clinical outcomes in terms of motor and non-motor effects and quality of life improvements~\cite{deuschl2013clinical, Kim2020clnm}. Neuromodulation has been tried with more acute conditions such as spinal cord injuries, where epidural electrical stimulation is applied to stimulate specific sensorimotor functions~\cite{Wenger2014clnm}. Closed-loop VNS has shown promising evidence of the prolonged effects in restoring neural circuitry with a study on rats~\cite{Ganzer2018clnm}. Less commonly, neurostimulation has also been explored with other conditions such as psychiatric illness~\cite{Lo2017clnm} and loss of control eating~\cite{Wu2020rns}. 

In a recent study, notable insights into neuromodulation advancements by 2035 are discussed~\cite{denison2022neuromodulation}. The study emphasizes the flexibility of neuromodulation techniques, enabling customization for precise brain regions or networks, which can be administered through different methods, including one-time treatment, continuous delivery, or in response to physiological changes. It emphasizes the need to understand brain networks and the mechanism of stimulation, as well as advancements in material science, miniaturization, energy storage, and delivery to expand the use of these devices.

\paragraph{\textbf{The neurophysiological information obtained from today's sensing-enabled devices is fueling the discovery of neurologic and psychiatric biomarkers}}
Treatment success in conditions like epilepsy and dementia is typically based on incomplete and potentially inaccurate patient reports, making treatment decisions challenging for neurologists. Electroencephalography (EEG) offers several advantages as a biomarker tool. It allows for monitoring brain activity with a high temporal and reasonable spatial resolution. ~\cite{meghdadi2022event,hawellek2022changes,lundstrom2021low,meghdadi2021resting,simonato2021identification}. One goal is to identify biomarkers that indicate physiological changes preceding the clinical effects of diseases, to assist neurologists in making treatment decisions. Biomarkers, such as beta frequencies in Parkinson's disease and seizure spikes, have already been identified for certain disorders. However, biomarkers can have their own limitations. For instance, high-frequency oscillations (HFOs), comprising fast and slow ripples (80-250~Hz, 250-500~Hz, respectively), are useful biomarkers for detecting epileptogenic tissue. However, conventional scalp-EEGs are limited in capturing HFOs due to their frequency sampling rate. Therefore, invasive procedures involving penetrating (e.g. Utah electrode array), depth (e.g. stereo-EEG), or brain-surface (e.g. electrocorticography, ECoG) electrodes are required to detect HFOs~\cite{simonato2021identification,sharifshazileh2021electronic}.

\paragraph{\textbf{Advanced neuromodulation techniques are needed}}

Non-invasive methods are advantageous as they do not require surgery, and patients prefer them. However, they may require a bulky wearable device or uncomfortable wearability features or otherwise pose risks of skin irritation and damage. This is particularly true regarding interfacing with the human head and brain. On the other hand, implantable devices can provide more spatial specificity and longer-term interaction and potentially require less patient interaction.

Non-invasive brain stimulation techniques include clinical techniques such as transcranial direct-current stimulation (tDCS) and transcranial magnetic stimulation (TMS). tDCS faces the challenge of avoiding skin irritation, while transcranial magnetic stimulation TMS uses high-current pulses to create magnetic fields that excite the nervous system. More advanced techniques, such as paired stimulation and temporal interference, may be applied to synchronize different brain regions and deliver therapy more deeply in the brain. Focused ultrasound is another technique being studied as a non-invasive means to stimulate the brain acoustically, but it requires a sizeable external transducer. Looking longer term, optogenetics may offer greater specificity in neural modulation by modifying neurons to express light-sensitive, allowing for the direct inhibition or excitation of activity in specific cell~\cite{katz2019vivo}. Advancements in electrophysiology and structural/functional imaging, such as functional magnetic resonance imaging (fMRI) and diffusion tensor imaging, as well as more precise and less invasive brain mappings techniques like TMS and stereotactic electrode implantation, will assist in the development of more sophisticated neural network models of neurological and psychiatric disorders.

Stimulation might also be adapted according to time-based biological rhythms. Researchers are now exploring how feed-forward mechanisms such as sleep–wake and other biological rhythms at multiple timescales might optimize device control. Multimodal approaches are also used in some cases, such as incorporating a heart rate sensor into some vagus nerve stimulation devices to activate stimulation when the heart rate exceeds a predetermined threshold since some seizures are associated with an acceleration in heart rate. 

Devices can also adjust stimulation on a pulse-per-pulse basis, and research is underway to modify responsive stimulation based on individual circadian and multi-day seizure rhythms to forecast times of greater seizure susceptibility.

\paragraph{\textbf{Future brain stimulation devices will use advanced algorithms that combine predictive models and responsive feedback mechanisms}}

In responsive neurostimulation, a neurostimulator device is surgically implanted within the patient's brain or near the affected area. This device has electrodes that constantly monitor the brain's electrical activity in real-time. It is programmed to detect abnormal electrical patterns or seizure onset based on predefined algorithms. One example is to treat epilepsy by continuously monitoring intracranial EEG and providing stimulation only when epileptiform activity is detected. NeuroPace developed one of the first responsive neurostimulation (RNS) systems for epilepsy that detects abnormal brain activity and responds in real-time~\cite{krucoff2021operative}. This reduces the amount of stimulation required and improves the accuracy of the treatment. This closed-loop system utilizes databases, modeling, and machine learning to enhance performance while gathering data necessitates additional telemetry and data storage.

Neuromodulation will increasingly depend on data science for better outcomes. However, interpreting brain data obtained chronically or in real-time requires advanced analytics that rely on deep learning algorithms and intensive computational capabilities, which are unsuitable for current hardware and software approaches for on-chip learning ~\cite{kremen2018integrating}.

\paragraph{\textbf{Can these devices be smarter, extraordinarily energy-efficient and perform truly real-time closed-loop therapy?}}

Neurotechnology research has seen a surge in startups and companies over the past decade, but on-chip computation is currently limited to simple signal processing and feature extraction. Existing systems such as the responsive neurostimulator~\cite{carrette2015responsive}, Percept PC~\cite{jimenez2021device} and Summit RC+S~\footnote{Some of these techniques are approved only for investigational use.}~\cite{stanslaski2018chronically} rely on external systems with advanced machine learning algorithms for accurate symptom tracking. However, this alternative also faces challenges such as increased power consumption due to wireless data telemetry and significant latency in the feedback loop (several hundred milliseconds) relative to potential latency for an on-device equivalent. Integrating this alternative approach could reduce the effectiveness of closed-loop stimulation and an increased need for more frequent battery replacements or recharges of implanted batteries. The goal is to design the next generation of intelligent neuromodulation systems with more on-chip computing, energy efficiency, and overall miniaturization~\cite {yoo2021neural}. 

\section{Neuromorphic Neuromodulation: Driving the next generation of on-device AI-revolution in electroceuticals}

In this perspective, we aim to explain why neuromorphic computing may represent a potential solution for making tiny and embedded devices in electroceuticals by offering advantages, challenges and feasible applications.

\paragraph{\textbf{Data telemetry is power hungry}}

The rapid advancement of AI and neural networks has led to computers exhibiting impressive cognitive abilities. However, the challenge remains in reducing the computational cost and achieving brain-like efficiency. Deep neural networks form the basis of state-of-the-art AI as it stands, and these networks rely on computing systems, from the transistors to hugely memory-intensive graphics processing units (GPUs), which consume substantial energy in their general-purpose and conventional computing architectures. 

Deep neural networks are trained on energy-intensive servers, resulting in remarkable accuracy but high energy consumption. For example, running a model on an intelligent glass-embedded processor would exhaust its battery (2.1~Wh) within a span of 25 minutes~\cite{venkataramani2016efficient}. Nonetheless, its high power consumption renders it unsuitable for bio-electronic medicine applications prioritising low energy usage. To achieve energy-efficient machine intelligence, there is potential in developing customized hardware that mimics the structure of biological brains to enable more local cognitive capability in processing, identifying and acting on personalised features. 

External data processing in implantable devices requires wireless data telemetry limited by bandwidth, communication range, interference, and, more importantly, energy requirement. If there is a real-time processing need, such external interaction would hinder timely reaction to signal features and potential efficacy issues. On the other hand, on-device (edge) computing solutions enable the immediate processing of recorded signals and facilitate closed-loop interventions. ~\cite{kiourti2012review,movassaghi2014wireless}. If we can expand this edge computing capability beyond inference-only to on-device learning, the capability of the device to be truly personalised is greatly enhanced, and its potential for efficacy improvements is increased.

\paragraph{\textbf{Can we take inspiration from the brain through neuromorphic?}}

The human brain possesses a remarkable computational power ranging from $10^{13}$ to $10^{16}$ operations per second, with a power consumption of approximately 20~W \footnote{This is an indicative figure based on whole body metabolic studies.}. In contrast, a computer performing a classification task requires around 250~W. The brain consists of billions of neurons ($\sim 9 \times 10^9$) connected by trillions of synapses ($\sim 3 \times 10^{14}$), allowing for information processing at a rate of approximately $6 \times 10^{16}$ bits per second~\cite{martins2019human,fares2022realm}.

\begin{figure*}
\centering
\includegraphics[width=0.90\textwidth]{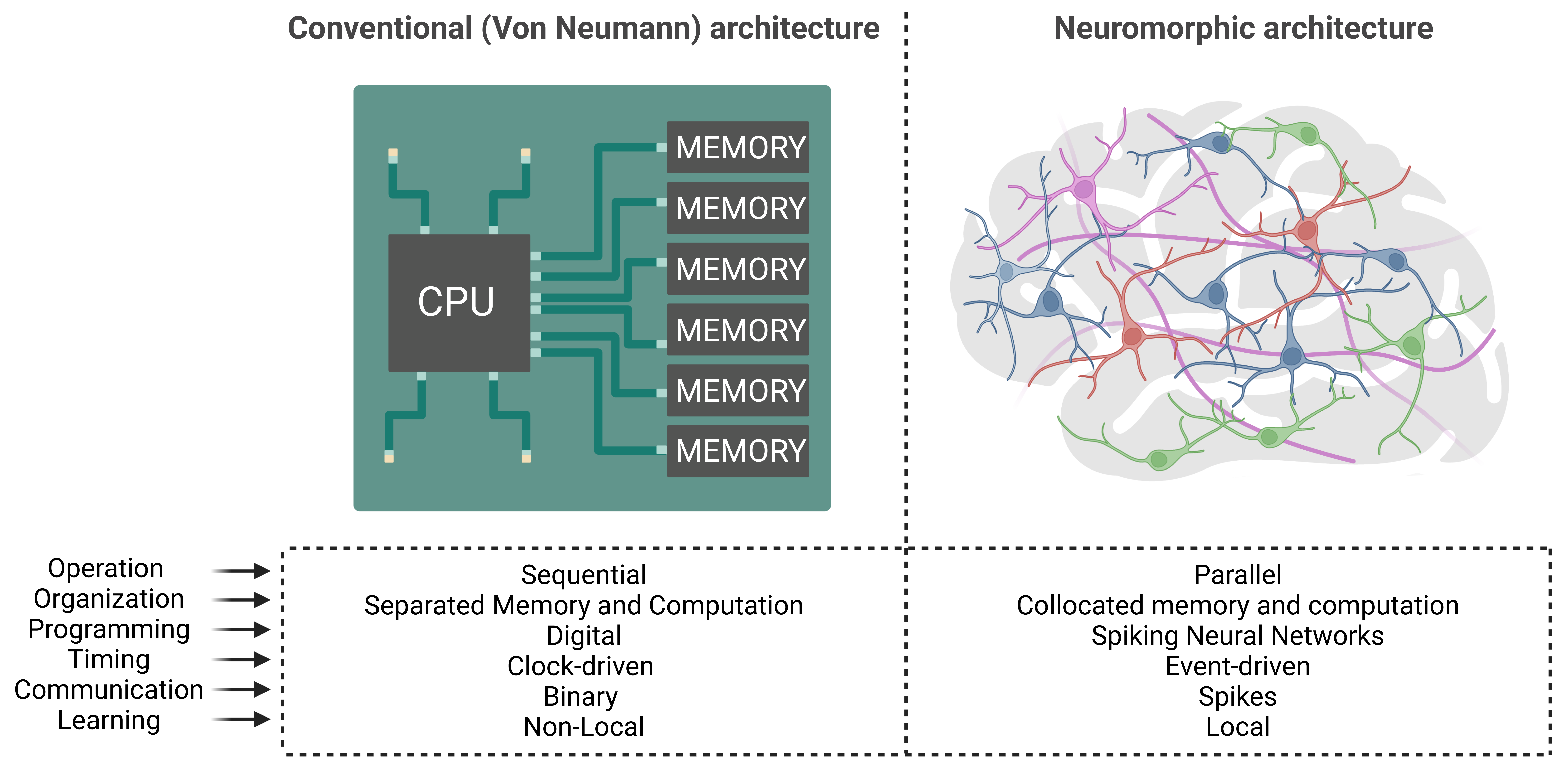}
\caption{Contrast between conventional (von Neumann, e.g. CPUs) architecture with bio-inspired (non-von Neumann, e.g. neuromorphic) architecture. Conventional computers rely on sequential, clock-driven (synchronous) binary operations, separating memory and computation units. In contrast, the human brain employs event-driven (asynchronous), neural action potentials (spikes), with a great network capacity for parallel processing and capability for local learning mechanisms. These basic, seemingly shallow, yet fundamental distinctions contribute to the brain's inherent superiority in terms of energy efficiency, positioning it as a promising avenue for custom or general-purpose integrated circuits and computing architectures development.}
\label{fig:neuromorphicomputing}
\end{figure*}

In a recent research investigation exploring the prospects of neuromodulation over a decade~\cite{afsaneh2023role}, the article discusses the potential of neuromorphic chips for implanted body-machine systems, which mimic the co-location of logic and memory, hyper-connectivity, and parallel processing of the human brain, as shown in Fig.~\ref{fig:neuromorphicomputing}.

The field of neuromorphic computing has seen significant advancements in industry and academia~\cite{fares2022realm}. Some notable industrial neuromorphic chips include IBM's TrueNorth and Intel's Loihi, each with specific software toolchains and applications. Several large-scale neuromorphic chips, such as BrainScales, SpiNNaker, NeuroGrid, IFAT, and DYNAPs, have been developed as part of the European Union Human Brain Project. These chips have succeeded in keyword spotting, medical image analysis, and object detection tasks. In contrast, research also focuses on creating general-purpose re-configurable neuromorphic platforms that bridge hardware and software frameworks for broader applications. One example is the Tianjic chip, which supports neuromorphic spiking neural networks (SNNs) and traditional artificial neural networks (ANNs). Spinnaker is a general-purpose accelerator that can handle various workloads, including those without spiking behavior. These large-scale neuromorphic chips are implemented in a mix of digital, analog, or mixed-signal configurations, depending on the specific functionalities required~\cite{Furber2014SpiNNaker,Davies2018Loihi,Mostafa2015EventBased,Merolla2014TrueNorth,Schemmel2010WaferScale, schemmel2021accelerated, Benjamin2014Neurogrid, Thakur2018LargeScale, Moradi2018Scalable, Blouw2019KeywordSpotting, Getty2021DeepMedical}. Table ~\ref{tablefirst} overviews some of the most prominent current neuromorphic chips. Notably, there is a wide range of neuromorphic chips, but we consider them mostly commercially available for demonstration. 

In another perspective~\cite{donati2023neuromorphic}, there is a discussion about the limitations of current devices for reading and stimulating peripheral nervous systems with current neuromodulation techniques and how neuromorphic circuits represent an ideal solution to improve bioelectric medicine. For instance, adaptive closed-loop systems using neuromorphic engineering can better control symptoms and side effects by continuously monitoring physiological signals and adapting in real-time. These systems use mixed-mode analog/digital transistors and consume ultra-low power. Neuromorphic engineering can overcome bandwidth and power consumption limitations, improving neural data acquisition and processing~\cite{chicca2014neuromorphic}. Analog neuromorphic front-ends offer a low-power solution to the challenges of high-bandwidth requirements and multiple-channel processing in neural recording. They directly process analog signals and convert them into spikes for downstream spiking neural network processing.

When working with neuromorphic technologies, a key aspect is a need to process signals in the form of spike, independently of the recorded signal is a sequence of spikes, voltage membrane, or local field potential. Spikes allow sparse and asynchronous communication, which propagates information based on their precise firing time with variable delays allowing spatio-temporal pattern discrimination~\cite{mainen1995reliability}.

\begin{table*}[htbp]
     \caption{A benchmark of neuromorphic chips. This list is not extensive but considers the breakthroughs~\cite{ivanov2022neuromorphic, mehonic2022brain,liu2021low,young2019review}.}
     \label{tablefirst}
        \rowcolors{2}{gray!12}{white}
        \resizebox{\linewidth}{!}{
        \begin{tabularx}{\textwidth}{ | Y{2.5cm} | Y{1.1cm} | Y{1.4cm} | Y{1.2cm} | Y{1.2cm} | Y{1.5cm} | Y{1.2cm} | Y{1.2cm} | Y{1.3cm} | Y{1.2cm} | }
            \hline
            \textbf{Features /Chips} & \textbf{CPU} & \textbf{TrueNorth} 
 ~\cite{akopyan2015truenorth,young2019review}& \textbf{Loihi2}  ~\cite{intel2021neuromorphic} & \textbf{Tianjic} 
 ~\cite{deng2020tianjic,pei2019towards} & \textbf{SpiNNaker}  ~\cite{Furber2014SpiNNaker,HBPNeuromorphicComputing} & \textbf{Brain ScaleS-2}  ~\cite{HBPNeuromorphicComputing,pehle2022brainscales,grubl2020verification} & \textbf{DYNAP SEL}  ~\cite{moradi2017scalable,rueckert2020update} & \textbf{Akida}  ~\cite{demler2019brainchip}   & \textbf{Human Brain}  ~\cite{martins2016human,martins2019human} \\
            \hline
            \hline
            
            \textbf{Event-driven} & No & Yes & Yes & Yes & No & Yes & Yes & Yes & Yes!\\
            
            \textbf{Analog or Digital} & Digital & Digital & Digital & Digital & Analog & Analog & Analog & Analog & Analog \\
            
            \textbf{Node (nm)} & 5 & 28 & 7 & 28 & 130 & 65 & 28 & 28 & - \\
            
            \textbf{In-memory computing} & No & Yes & Yes & Yes & Yes & Yes & Yes & Yes & Yes \\
            
            \textbf{Neurons / Synapses count} & - & 1M/256M & 1M/120M & 40K/10M & 16k/8M per chip & - & 1K/78K & 1.2M/10B & 20B/200T \\
            
            \textbf{Cores} & Vary & 4096 & 128 & 156 & 18 per chip & ~8 & 5 & 80 & - \\
            
            \textbf{Neurons / Synapses per Core} & - & 256/64K & 8K/900K & 256/64K & 800/1M & 512/130K & 256/ & 15K/125M & - \\
            
            \textbf{On-device learning} & Backprop /STDP & No & STDP /Surrogate backprop & No & STDP & STDP, R-STDP, homeostatic plasticity & STDP & Few-shot training & Diverse learning \\
            
            \textbf{Network compatibility} & ANN & SNN & SNN & ANN/SNN (hybrid) & Hybrid & SNN & SNN & ANN, SNN & SNN \\
            
            \textbf{Key properties} & Limited to on-chip training & First commercial neuromorphic chip & Fully programmable chip & First hybrid ANN/SNN & Scalable SNN simulation & Hybrid Plasticity & One plastic core & Continuous and few-shot learning & - \\

             \textbf{ASIC /FPGAs}  & - & ASIC & ASIC & ASIC & ASIC & Both ~\cite{stradmann2022demonstrating} & FPGAs & ASIC & - \\

             \textbf{GSOPS/W} & - & 400  & \(\sim 10\times{\rm Loihi1}^{\dagger}\) & 649-1278 ~\cite{deng2020tianjic,basu2022spiking}& 0.033 & \(>10\)& 33 & Akida & - \\
             
             \textbf{Memory type of store synaptic weights} & - & SRAM/ DRAM  & SRAM /DRAM & SRAM & SRAM & SRAM & SRAM & SRAM & Different brain regions\\
             
             \textbf{Routing schemes} & - & Grid/2D Mesh  & Grid/2D Mesh & 2D Mesh & Grid/2D Mesh & Hierarchical Tree & Hierarchical Tree & Grid/2D Mesh & Stochastic \\

             \hline
        \end{tabularx}
        }
    \begin{tablenotes}
      \small
      \item $^{\dagger}$~There is no direct data available but it is stated to be 10 times more than Loihi1.
    \end{tablenotes}
\end{table*}

Recent advances in neuromorphic computing provides an unconventional way (non-von Neumann architecture) to implement neural networks, which has made on-chip training of neural networks possible with very low power consumption and small footprint device~\cite{frenkel2020onchiptrain} performs on-chip training on a 32-mm$^2$ area of silicon with an accuracy of 95.3\% (as opposed to 97.5\% if off-chip training) tested with the MNIST dataset. Existing neuromorphic chips, such as the well-known IBM TrueNorth and Intel Loihi chips, are general-purpose chips that support various types of networks and configurable parameters (i.e., number of layers, kernel sizes, etc). However, their versatility comes with a cost of higher power consumption and heat dissipation. For implementation of neural networks with learning capability, a neuromorphic chip should be fully optimized for one specific application if continuous active learning is to be coupled with a medical device, especially implants that have strict constraints on temperature.

\paragraph{\textbf{On-device learning}}

AI systems utilizing ASIC (application-specific integrated circuit) with parallel multiply-and-accumulate (MAC) exhibit better inference and energy efficiency than GPUS. However, by performing MAC, the need for intensive data transfer between the MAC units and data buffers stills limits energy efficiency and therefore, are restricted to functioning solely in inference mode, whereas the human brain has the remarkable ability to learn continuously. Consequently, on-device learning emerges as a significant characteristic of neuromorphic systems. On-chip learning is indispensable for tailoring and personalizing smart devices to cater to individual user requirements. Moreover, it bolsters privacy by eliminating the need to transmit user data to the cloud ~\cite{zhang2020neuro,frenkel2021bottom,ivanov2022neuromorphic}.

Data sharing is crucial to attain real innovations and values, and organizations related to healthcare must follow standards related to the healthcare domain~\cite{savage2020doctors}. To protect individual privacy, fully anonimized or de-identified data can be shared according to criteria set by privacy and security standards such as the Health Insurance Portability and Accountability Act (HIPAA) in the USA~\cite{white2022data}.

In~\cite{surianarayanan2023convergence}, the article discusses the limitations of using artificial intelligence (AI) models in clinical applications, citing the misalignment of performance metrics used in the algorithms with clinical applications. To overcome this, AI models must be validated with prospective data and assessed in real-world setting. 

\subsection{Feasible application for neuromorphic neuromodulation}

\begin{figure*}
\centering
\includegraphics[width=0.9\textwidth]{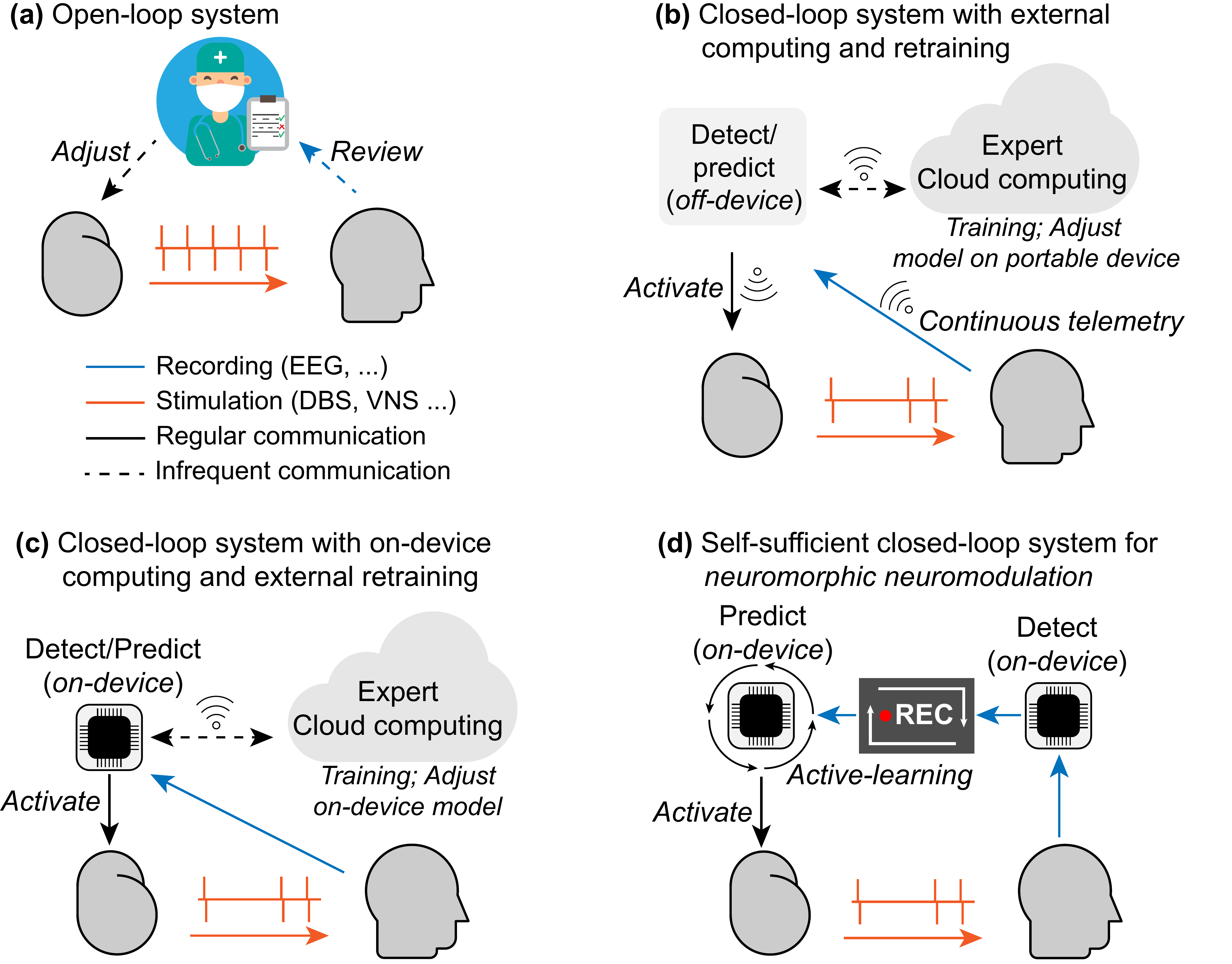}

\caption{Neuromodulation approaches. (a) An open-loop system with an expert who occasionally reviews the effectiveness of the system and adjusts the stimulation parameters accordingly. Such a system employs cyclic stimulation regardless of the current state of the target (e.g., brain state). (b) A closed-loop system with external computing for accessing the state of the target to condition the stimulation. The external computing component can be in the form of a portable device, e.g., tablet, or a local computer. The recording device continuously streams data (e.g., EEG signals) to the external computing where trained algorithms are executed to determine the target's state. The deployed algorithms on the external computing component can be updated occasionally by involving a review from an expert(s) and big data/cloud computing (retraining). (c) The computing component is embedded within the device, which eliminates the need for {\it continuous} streaming of data to the outside world~\cite{harrer2020seizure, pepin2021neuromodulation}. However, the on-device algorithms need to be occasionally updated to reflect the change in physiological signals (e.g., change of seizure patterns in epileptic patients). The device must also have sufficient memory to store the signals for the expert(s) to review and for the retraining that takes place in the cloud. (d) A neuromorphic neuromodulation system where the medical device can run and retrain its algorithm by itself without relying on external computing resources. The system utilizes labels generated by a detection model that has performance on-par with a human expert~\cite{Hannun2019cardiolevel, Jing2020expertSZdet, Scheuer2021szdetexpert} and a loop recorder to train a prediction model. The rapid improvements in neuromorphic computing~\cite{park2019a65nm,bohnstingl2019neuromorphic} have made on-device active learning possible.}
\label{rns:fig:overview}
\end{figure*}

Fig.~\ref{rns:fig:overview} demonstrates three main categories of neuromodulation devices, including those which are commercially available (a-b) and under investigation (c). In an open-loop system (Fig.~\ref{rns:fig:overview}(a)), stimulation parameters such as amplitude, frequency, and duty cycle are pre-determined by a clinician. Stimulation persists unless manually turned off by the patient or the clinician. The device can be re-programmed during a subsequent patient visit if the stimulation does not show effectiveness. In contrast, a closed-loop system activates the stimulation adaptive to physiological changes. The system in Fig.~\ref{rns:fig:overview} (b) continuously senses the patient's state (e.g., EEG signals) and streams it wirelessly to a portable device (or a bedside computer) that is in charge of analyzing the signals by using threshold-based rules or machine learning models, and turning it into a control signal to the stimulation~\cite{giftakis2014seizure, denison2017seizure, snyder2008methods, chen2021role, sladky2021distributed, stanslaski2018chronically, kremen2018integrating, mivalt2021electrical, stirling2021seizurefor}. For example, in the case of epilepsy, if the portable device detects a seizure onset or predicts an incoming seizure, then activates the stimulation. 

The closed-loop system with external computing resources reduces unnecessary stimulation compared to conventional systems. However, the system is exposed to a major downside. The high demand of power consumption for continuous streaming of data to external computing resources quickly reduces battery life or increases the size of the device. Wireless communication between the device and the external computing resource is prone to complications such as lost/jammed connection, and security risks with interference. Fig.~\ref{rns:fig:overview}(c) illustrates how an alternative closed-loop system can address the complications from continuous data stream of standard closed-loop systems by incorporating an on-device computing unit to the neuro-modulation device without reliance on external computational power to host the control algorithms~\cite{Rhew2015selfdbs, harrer2020seizure, pepin2021neuromodulation}. In these closed-loop system models (Fig.~\ref{rns:fig:overview}(b) and (c)), the optimization of models (on-device or off-device) must still be performed regularly and involves a human expert and/or cloud computation to adapt with the changes in the patient's conditions and/or in the underlying disease~\cite{sun2014closed}. This implies that the patient's data needs to be stored in the external device (Fig.~\ref{rns:fig:overview}(b)) or in the implantable device (Fig.~\ref{rns:fig:overview}(c)) and regularly be uploaded to the cloud. The data will also need to be analyzed or labeled by a human expert so it can be used to update threshold-based rules or to retrain the machine learning models.

It should be noted that none of the aforementioned methods offer on-device training and re-training, and requires expert involvement for regular retraining~\cite{ashourvan2020model}, which limits the scalability of the system to a large number of patients~\cite{Karuppiah2018active}. The idea of on-device active learning proposed in~\cite{xiao2017adaptive} relied on an ideal detection and deterministic feature extraction technique to actively train a prediction model without expert intervention or external computational resources. However, we argue that deterministic feature extraction may lose its efficacy over time because the underlying disease is evolving. 

Our alternative is neuromorphic neuromodulation, a computationally self-sufficient closed-loop system, shown in Fig.~\ref{rns:fig:overview}(d). Our proposed system eliminates the requirements of continuous data telemetry and the reliance on external computational resources. We believe our self-contained system can provide an ultimate personalized closed-loop neuromodulation system. The vision is ambitious but not impractical. In the following sections, we explain in detail how we can accomplish such a system.

\paragraph{\textbf{Physiological event detection becoming more trustworthy}} There has been an increasing body of work using automatic annotations of physiological data, and recently, they are approaching human expert level. Some examples include arrhythmia detection and classification~\cite{Hannun2019cardiolevel} detection of epileptiform discharges,~\cite{Jing2020expertSZdet} and seizure marking~\cite{Scheuer2021szdetexpert}. These advancements are not only helpful in automatic health monitoring with a reduced burden for clinicians, but also pave a new way to exploit unlabeled data which is easier to collect. For instance, generated labels by algorithms can be used to train other predictive algorithms without human expert intervention. This leads to the birth of AURA, an Adaptive, Unlabeled and Real-time Approximate-Learning platform ~\cite{yang2022weak}

As an example, the success rate of detecting a seizure either during or right after it starts is significantly higher than attempting to forecast seizures beforehand~\cite{mormann2007seizure}. Interactival activities are a promising biomarker for forecasting epileptic seizures~\cite{yang2022weak} 

\paragraph{\textbf{Continuous Learning with AURA}} Continuous Learning with AURA. At the core of AURA, a high performance physiological event (e.g., epileptic seizure) detection model acts as an algorithmic “human expert” to generate labels on-the-fly as the signal arrives. The generated labels are paired with recorded signals from a loop recorder to be used as a training dataset for a predictive model (e.g., seizure forecasting). It is worth noting that while the detection model or label generator must have high performance, it does not necessarily need to be perfect. In fact, imperfect labeling from a mix of clinicians and medical students with varied levels of experience has shown to remain effective in training a deep learning model to perform seizure detection at a high level of accuracy and generalization. However, some limitations with AURA is that is limited by a buffer for continuous learning and relies on backpropagation for training. 

\paragraph{\textbf{Embrace multi-modal signals}} As part of physiological monitoring, it is usual that there are multiple signals being recorded. Combining signals from multiple sources has the potential to improve the performance of a detection/prediction model~\cite{greene2007combination, valderrama2010patient}. It is important to note that depending on performance requirements, power consumption and/or heat dissipation, the detection model may use a different set of sensory modalities from the prediction counterpart.

\paragraph{\textbf{Back-propagation: implausible biological ways and issues with Neuromorphic hardware}} Artificial neural networks utilize the back-propagation algorithm and gradient descent optimization to adjust the synaptic weights between neurons. By overwriting all network parameters between tasks, this contributes to catastrophic forgetting. Back-propagation also suffers from weight-symmetry problem~\cite{liao2016important}, freezing of neural activity~\cite{whittington2019theories}, and non-local weight update~\cite{lillicrap2020backpropagation}. Back-propagation also face problem when dealing with adversarial attacks~\cite{akrout2019adversarial} where incoming noisy data may arrive. Although back-propagation is efficient, its implementation in analog VLSI requires excessive computational hardware~\cite{jabri1992weight}

A novel solution relies in the forward-forward algorithm, taking inspiration of how our brains neurons deals with information and help mitigate the above mention problems.~\cite{dellaferrera2022error,hinton2022forward}. The forward-forward algorithm has thus far been demonstrated solely on static datasets like CIFAR-10, implying that its current scope is somewhat restricted. Nevertheless, acknowledging its potential is certainly intriguing.

\paragraph{\textbf{Metaplastic neural models for continuous learning}}

In computational neuroscience, a notable conundrum is the brain's ability to acquire novel memories and skills efficiently while steadfastly maintaining existing ones over a lifetime.~\cite{mermillod2013stability}

Synaptic plasticity refers to the ability of neuronal connections to alter their strength over time, is widely regarded as the underlying physical mechanism of learning in the brain. Knowledge is believed to be spread out throughout neuronal networks, with individual synapses holding multiple memories.  Given this shared nature of memory storage, synapses must exhibit both flexibility in response to new experiences and stability to retain existing ones. This paradox is well know as the stability-plasticity dilemma~\cite{mermillod2013stability}.

A novel, bio-inspired strategy to mitigate catastrophic forgetting and to continue learning, is synaptic meta-plasticity\cite{laborieux2021synaptic}. Based on neuroscience studies, it has been formulated that synapses have the ability to adjust their plasticity. The authors utilized a binarized neural network that can be implemented on neuromorphic platforms. Each synapse includes a hidden weight that is learned during training and a binary weight that is utilized during inference. To address catastrophic forgetting in multi-task learning, the authors introduced a meta-learning method that modulates the hidden weights via a function \textit{fmeta(Wh)}. This modulation is applied to negative updates of positive hidden weights and positive updates of negative hidden weights. Future studies will be inclined towards metaplastic behavior with SNN compabilities.  

\paragraph{\textbf{Smart Algorithms and the role of Neuroscience as inspiration}}

A study focuses on early seizure detection for closed-loop direct neurostimulation devices in epilepsy with neuromorphic chips. It discusses the successful transfer of a convolutional neural network (CNN) to the neuromorphic TrueNorth chip. The CNN implementation on the chip demonstrates favorable detection accuracy, low memory requirements, and efficient runtime cycles. However, it acknowledges that CNNs still rely on backpropagation, which can lead to issues such as catastrophic forgetting and increased computational expense~\cite{dumpelmann2019early,esser2016cover,kiral2018epileptic}

A study introduces a real-time seizure prediction system for epilepsy, utilizing deep learning and big data. The system is designed to run on a low-power neuromorphic chip and can be integrated into a wearable device. Authors suggests applications in closed-loop therapeutic devices for epilepsy treatment.~\cite{kiral2018epileptic}

Classical pattern recognition that predates machine learning relies on manually extracted features and has limited processing capability. ANN (Artificial Neural Networks), inspired by the brain's neural network, enables automatic feature discovery. RNN and LSTM incorporate working memory for sequential data processing. CNN, inspired by the brain's visual stream, extracts complex features for object recognition. RL learns through rewards and punishments, mimicking animal conditioning. Deep RL combines deep neural networks for complex environments. CLS theory and hippocampal learning inspire deep reinforcement learning. As discussed previously, this models are yet far to achieve energy-efficient by themselves. 

\paragraph{\textbf{Spiking Neural Networks: the third generation of deep neural networks}} Spiking Neural Networks (SNNs) provide an alternative approach by mimicking the behavior of biological neurons and offering potential energy efficiency advantages, making them suitable for resource-constrained environments like edge devices~\cite{surianarayanan2023convergence}. 

Training approaches for Spiking Neural Networks (SNNs) can be categorized into different methods. One common approach is to directly train the SNN itself using spike-based quasi-backpropagation. This involves adjusting the weights and connections of the network based on spike activity. Another approach involves training a traditional Artificial Neural Network (ANN) and then mapping it into an SNN. The ANN is trained using conventional techniques, and the resulting trained weights and connections are then transferred to the SNN. Reservoir computing is another technique used in SNNs, where the network is structured with an input layer, a reservoir layer, and a readout layer. The reservoir acts as a dynamic memory, and the readout layer is trained to interpret the reservoir's activity. Evolutionary approaches are also utilized, where the structures and parameters of the SNN evolve over time through genetic algorithms or other evolutionary strategies. Lastly, spike-timing-dependent plasticity is an approach based on the synaptic plasticity mechanism, where the weights of the connections are adjusted according to the relative spike timings between pre and post-synaptic neurons. These training approaches provide different strategies for training SNNs, each with its own advantages and applications.~\cite{schuman2022opportunities, mead1990neuromorphic, mead2020we,lee2016training,kulkarni2018spiking,tanaka2019recent,stockl2021optimized,schliebs2013evolving}

It is worth noting that training a conventional DNN and then mapping it to neuromorphic hardware, especially emerging hardware systems, can result in a reduction in accuracy not only because of the change from DNNs to SNNs, but also because of the neuromorphic hardware itself. By solely relying on supervised learning, we eliminate the potential for adapting to new types of data. Exploring alternative approaches is crucial for pushing the boundaries of machine learning. 

\paragraph{\textbf{Liquid Time Constant Based Models}}

Liquid Time Neural Networks are a class of time-continuous recurrent neural networks models that posses stable and bounded behaviour, improving performance on time-series prediction tasks. Their low complexity allow for a better representation of the hidden states, and adapting to changing conditions such as autonomous driving and medical time-series data~\cite{hasani2020liquid,lechner2020neural,hasani2022liquid,hasani2022closed,chahine2023robust,beveridge2022interpretable,banerjee2009descriptive}

We used Liquid-Time constant in different scenarios such as models on shallow bio-inspired models and models that do not need preprocessing.

Neural circuit policies (NCP)~\cite{hasani2022closed} based on biological inspiration, have been shown to possess greater generality, interpretability, and robustness, despite their simplicity, in comparison to significantly larger black-box learning systems. 

\begin{figure*}
\centering
\includegraphics[width=0.9\textwidth]{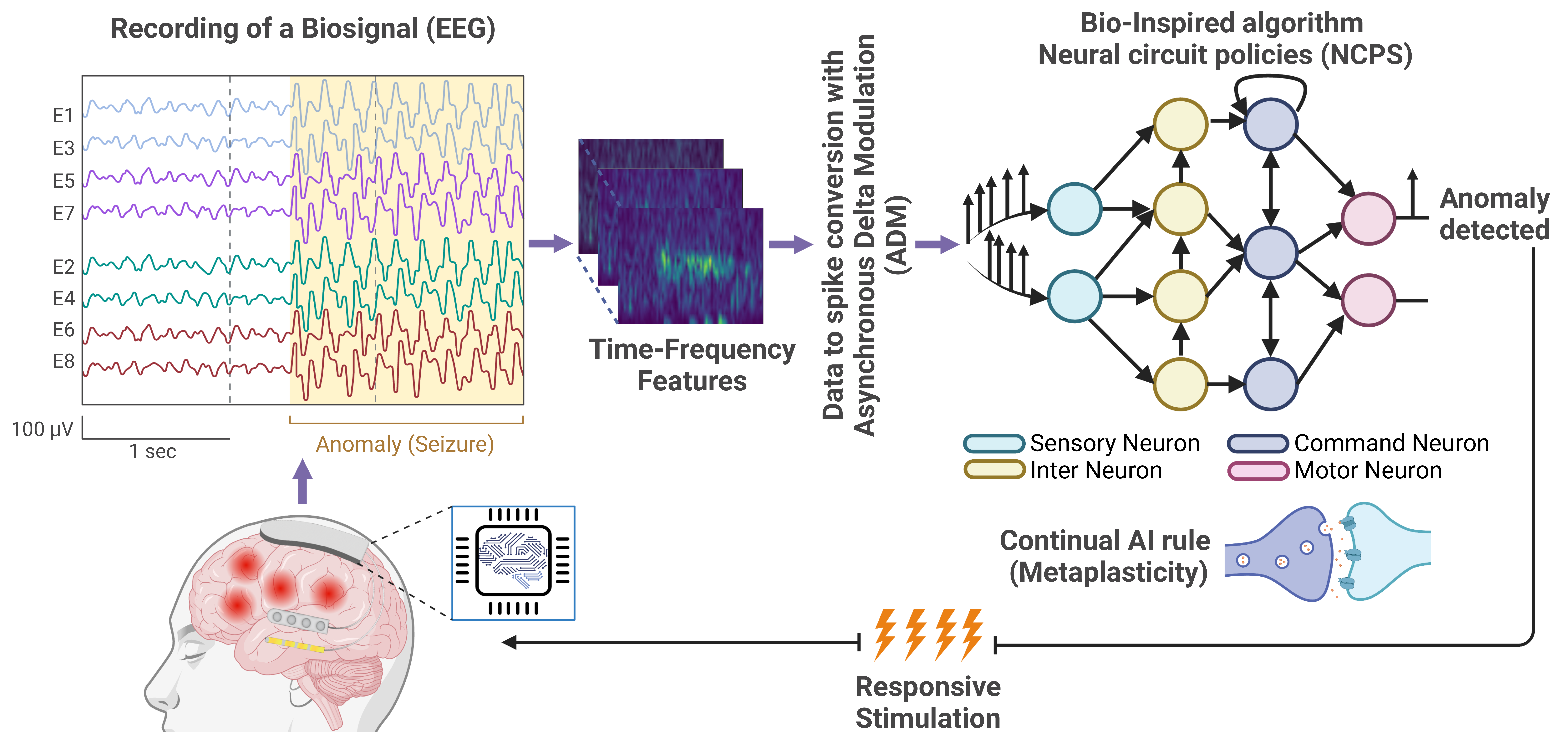}
\caption{{Neuromorphic neuromodulation employing bio-inspired learning rules represents a cutting-edge paradigm in the field of neural systems. This innovative approach enables the development of on-device learning capabilities, thereby facilitating the seamless integration and real-time processing of continuous bio-signals. By leveraging this neuromorphic system, it becomes possible to dynamically adapt to extracted features, subsequently converting them into spikes. These spikes are then fed into a shallow, sparse, and bio-inspired algorithm that utilizes a continuous learning rule for precise adjustment, ultimately yielding responsive stimulation tailored to each individual patient. This approach eliminates the reliance on cloud computing, ensuring a self-contained and autonomous system}}
\label{rns:fig:neuralcircuitpolicies}
\end{figure*}

We tested a CNN-NPC model to see robustness in raw EEG data but results exhibited sub-optimal performance. Our assertion is that the model does not possess the capability to effectively analyze or represent raw time series data. For instance, this models still depends on feature engineering in preprocessing like STFT to capture clinical biomarker for a neuromorphic neuromodulation system with bio-inspired architecture, as presented in Fig. ~\ref{rns:fig:neuralcircuitpolicies}

In this way, we used Liquid-S4. This model is a combination between liquid-time constant and S4 structure state-space models. S4 has the ability model time series data such as ECG and EEG with high-accuracy and fidelity without need of preprocessing~\cite{delorme2023eeg,tang2023modeling}. We run preliminary test results with raw EEG data from TUH and CHB-MIT. We then tested in sample in each dataset to evaluate if the model can outperform previous models that used preprocessing as STFT, ICA. Results are show in Table˜\ref{tab:liquid-s4}

\begin{table}[ht]
  \centering
  \caption{Liquid-S4: Global average tested results in TUH and CHB-MIT datasets.}
  \label{tab:liquid-s4}
  {\fontsize{10}{11}\selectfont  
  \begin{tabular}{c | c | c | c | c}
    {\bf Datasets} & {\bf AUROC} & {\bf AUPRC} & {\bf Precision} & {\bf Sensitivity} \\
    \hline
    \hline
    {\bf CHB-MIT} & 0.99 & 0.86 & 0.87 & 0.81 \\
    \hline
    {\bf TUH} & 0.89 & 0.77 & 0.78 & 0.60  \\
  \end{tabular}
  }
\end{table}

\begin{table}[ht]
  \centering
  \caption{Parameters used in the experiment with Liquid-S4, showing efficacy and promising of liquid time constant in S4 models.}
  \label{tab:paramused}
  {\fontsize{10}{11}\selectfont \begin{tabular}{l|c}
    Layers used & 1 - 4\\ \hline
    Number of parameters & 70K - 300K \\ \hline
    Memory size & 0.8 MB - 4.0 MB \\ \hline
    Liquid degree\textsuperscript{*} & 2 - 5\\
  \end{tabular}
  }
    \begin{tablenotes}
        \item[] {\textsuperscript{*}  Liquid degree refers as a implemented kernel to s4 from \textit {{2 to N}}}
    \end{tablenotes}
\end{table}

Despite their robustness to handle raw data, their implementation on neuromorphic devices remains a challenge. 

A proposed concept of integration of Liquid Time Constant with Spiking Neural Networks~\cite{bidollahkhani2023ltc} enable simple models compatible with embedded neuromorphic devices

In a recent study, there is a new approach called forward-propagation through time (FPTT) learning, which incorporates liquid time-constant spiking neurons. This method addresses the limitations associated with networks of complex spiking neurons and enables online learning of exceptionally long sequences~\cite{yin2023accurate}. This paves the possibility of using the liquid time constant with neuromorphic strategies for dealing with multimodal biosignals. We envision that this suitable, bio-inspired models may be reliable and power-friendly. 

\paragraph{\textbf{Risk of false alarms or unnecessary stimulation}} There has been a paradigm shift from open-loop stimulation (e.g. periodic DBS) to adaptive or closed-loop stimulation. The idea of adaptive neuromodulation was sparked in the early 2000s and has been quickly adopted over recent years. Perhaps the biggest concern related to responsive stimulation is how to guarantee stimulation is triggered when and only when required, i.e., high sensitivity and low false positives. Based on research findings showing that chronic brain stimulation can be performed safely with appropriate control of charge density \cite{gordon1990parameters, Morrell2006stim}, we can allow the stimulation activation system to have as high a sensitivity as possible with an increased number of false positives as a trade-off. A responsive stimulation with many false positives can be considered equivalent to an open-loop system that performs cyclic stimulation, given that there is a control of charge density and stimulation frequency/duration in place.

\paragraph{\textbf{Elimination of continuous data communication}} Wireless data communication can consume half or more of the total power consumption of the whole EEG recording implant~\cite{Yin2013implant, sawan2013wireless}. Neuralink reduces the frequency of sending data outside to every 25 ms and places a rechargeable battery and an inductive charger in the implant~\cite{Neuralink2021}. There is no doubt that continuous data communication is critical for disease diagnosis or a brain-computer interface. It is also inevitable for responsive closed-loop neurostimulation systems where some computation (training of the event detection/prediction model) needs to be performed with an external device or on the cloud~\cite{chen2021role}. The event detection/prediction model of such a system needs to regularly be re-trained with the most recent historical data to adapt to the change in physiological signals of the patients. The updated model will then be off-loaded to the implant. This procedure generates inter-dependencies among data telemetry, model training on external hardware, and off-loading of the updated model, where any point of failure would impact the efficacy of the stimulation. We argue that if the implant can efficiently learn from the data by itself to adapt to the change in physiological signals and become patient-specific, data communication to the outside world can be eliminated. The system does not send data out except for debugging purposes. In our vision, the system acts like interventional medicine because it is prescribed as a medical device for treatment based on diagnostics, where the device will work on its own to suppress or remedy the condition or disease on an as-needed basis.

\paragraph{\textbf{Estimation of power consumption of a fully integrated system}}

\begin{figure}
\centering
\includegraphics[width=0.53\textwidth]{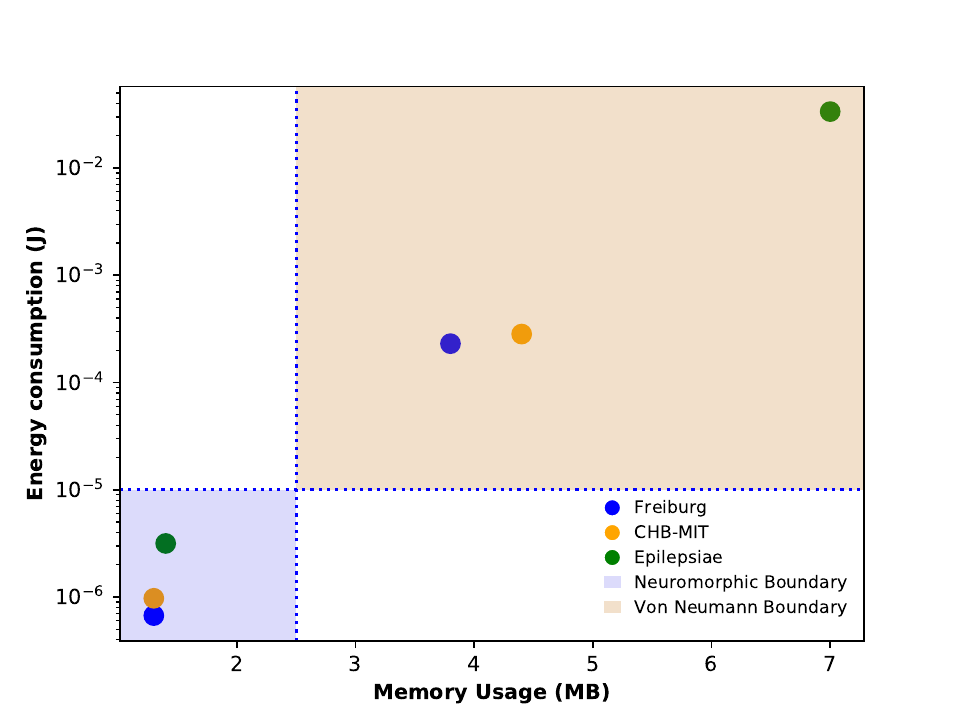}
\caption{Memory usage (a) and energy consumption (b) achieved by a conventional AI model (ConvLSTM) vs a spiking neural network (SNN) model during testing}
\label{fig:spikingenergyconsumption}
\end{figure}

A study conducted with spiking neural networks on EEG datasets (Freiburg, CHB-MIT, Epilepsiae) for seizure detection was proposed by~\cite{yang2023neuromorphic} where they demonstrate the capabilities of neuromorphic approaches to reduce the memory usage and energy consumption from ten to thousands of magnitude in comparison like running in conventional GPUs devices with conventional AI algorithms. The results of this study are demonstrated in Fig. ~\ref{fig:spikingenergyconsumption}

That study served as inspiration for estimating the power consumption of a fully integrated system. 

We assume the input signal (e.g., EEG) has 10 channels and a sampling rate of 128 Hz. For the sake of simplicity, both detection and prediction networks use similar convolutional long short-term memory (ConvLSTM) network architectures proposed in~\cite{yang2021continental}, which consists of three ConvLSTM layers followed by two fully-connected layers. The detection and prediction algorithms use input windows of 10 and 30 seconds, respectively. The inputs are divided into 50\% overlapping 1-second sub-windows to be fed to a ConvLSTM network. Using the Loihi neuromorphic chip as a reference and a real-time batch size of 1 (one input is processed at a time), the cost of inference for a single data sample is 25~mJ and 77~mJ for the detection and prediction algorithms, respectively. Since detection occurs every 10 seconds and prediction every 30 seconds, the energy can be amortized over time, with the total power consumption for inference of both networks being (2.6~mW$+$2.5~mW) or 5.1~mW. Note that this power consumption can be reduced if inference is parallelized into batches and distributed across a longer time interval. For example, with a batch size 32, the total inference power consumption becomes 160~$\mu$W.

Regarding the training of the prediction network, as the architecture is fixed, the backward pass can be completed simultaneously with the forward pass using a deterministic mode of forward-mode auto-differentiation; therefore, the gradient calculation cost is similar to the inference cost. However, there is the additional cost of routing gradients and using them to update weights. Given the network has 31.5M parameters, and the energy for updating each weight is 120~pJ (Loihi), the total cost of weight update is 3.78~mJ. The total required energy for training the prediction algorithm is (77~mJ$+$3.78~mJ) or 81.78~mJ. This training step occurs every 30 seconds, so its power consumption is 2.73~mW (batch size$=$1) or 85~$\mu$W (batch size$=$32). 

\begin{figure}[!ht]
\centering
\includegraphics[width=0.45\textwidth]{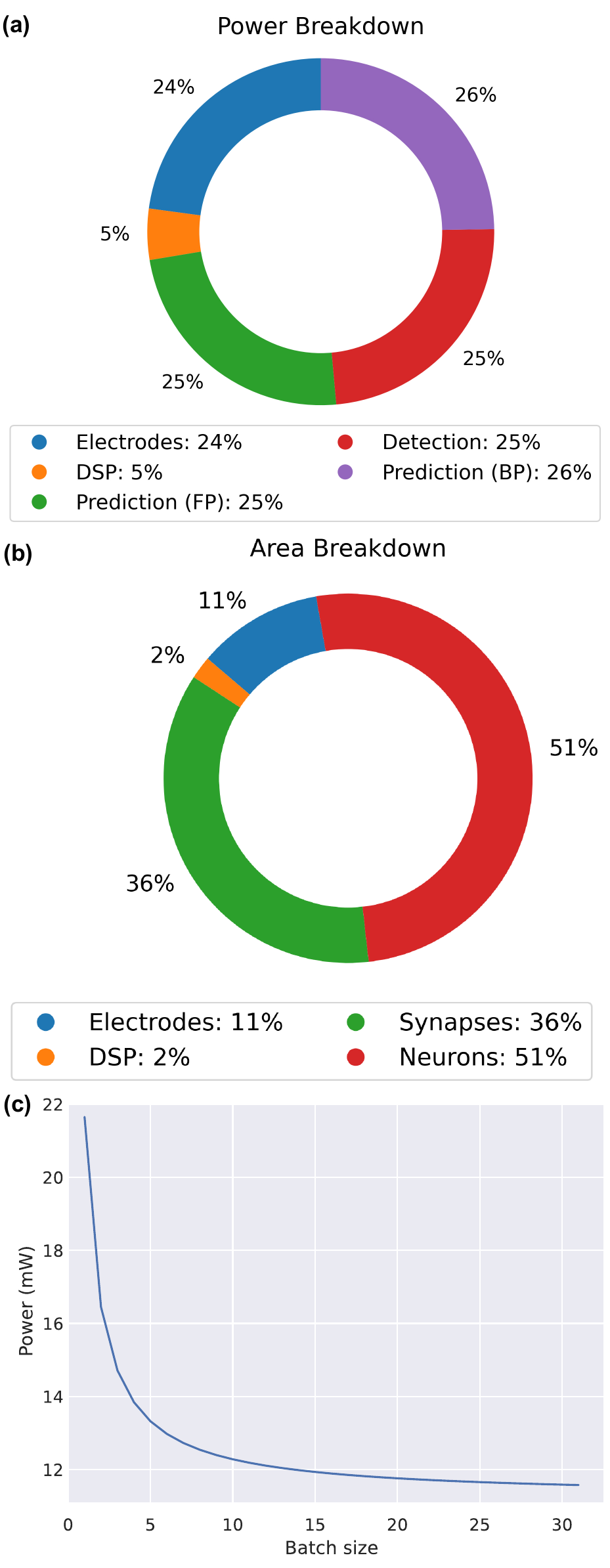}
     \caption{A system breakdown of on-board neuromorphic neuromodulation. (a) Area breakdown (DSP: digital signal processing blocks). (b) Power breakdown (FP: forward pass, BP: backward pass). (c) As batch size increases, the update interval decreases, leading to a reduction in power.}
     \label{fig:breakdown}
\end{figure}

With a custom design, the EEG front-end's power consumption can be optimized to less than 30~$\mu$W~\cite{Valle2016EEGlow}. Using a commercially available rechargeable Li-Po battery with a size 2.7$\times$30$\times$34~mm and a capacity of 240~mAh at 3.7V~\cite{LiPolBatteryCo2021}, the whole system can be powered for at least 24 hours before a recharge. The battery life can be considerably extended with calculations in batches, with the only trade-off of a slight delay in obtaining results. Considering the size of the detection and prediction networks and adhering to the synaptic density of the TrueNorth chip, we estimate a required area of $\sim$62~mm$^2$ to implement the whole system. A power and area breakdown are provided in Fig.~\ref{fig:breakdown}. The total system for 1 batch size can be up to 22~mW, where the inference of both networks of prediction and detection are consuming 5.1~mW; this represents around the 50\% of the system, while the training of the prediction network can be around 5.3~mW, which takes 27\% of the power of the whole system. Electrodes can consume as 5~mW~\cite{tang2020EEG}

With a digital signal processing block mostly related to doing fast Fourier transform (FFT) to extract essential features, 27.72~nJ per 128 FFT can be consumed. Considering a 50\% overlap, 10 channels, and 2 networks, it can take as 1.1~mW, representing only a 5\% of the power \cite{noor2018design}

\paragraph{\textbf{Challenges and opportunities}}

To this end, we will need a better mapping of the neural circuits associated with the treated pathophysiology; at the signal level, we will need better decoders of the neural language associated with the pathophysiological states and more precise therapeutic patterns of electrical impulses targeting the rate, even the timing of spikes. Generating such adaptive and precise neuromodulators will require a multidisciplinary effort: the development of neuromorphic circuits for real-time spike processing will translate the biological understanding of what is happening at the neural level in health and disease~\cite{michmizos2017computational}

A major challenge in neuromorphic algorithmic development is the absence of established benchmarks, metrics, and challenge problems. Without standardized evaluation criteria, it becomes highly challenging to compare and assess the suitability of different hardware systems for specific algorithms or applications. 

Neuromorphic hardware development involves extensive research into new materials and devices. Materials are a consideration for developing neuromorphic chips for neuromodulation.\cite{minnekhanov2019parylene}. The use of nanomaterials such as carbon-based nanostructures is suggested for generating bio-compatible probes for these systems. On the other hand, FDA-approved biocompatible and soft polymeric material, parylene, is shown to be used for neuromorphic building block development~\cite{sangwan2020neuromorphic}. Thus the development of memristive systems based on it provides prospects for the hardware realization of artificial neural networks for wearable and biomedical applications.

There are opportunities in neuromorphic for a software-hardware co-design process, where algorithms and applications influence the underlying hardware design. This approach allows for customization of the hardware implementation to meet specific application requirements. It also encourages exploring beyond digital computing and considering analogue and mixed-signal computing, mirroring the stochastic nature of biological neural computation. This co-design process presents exciting prospects for advancing neuromorphic systems and reimagining the computing stack.

Nanowires networks (NWNs) are a promising hardware approach to mimic the physical nature of the brain such as the neurons and synapses. In~\cite{loeffler2023neuromorphic}, they demonstrate how their results are analogous to synaptic metaplasticity for consolidating memory via strengthed synaptic pathways. Neuromorphic systems that learn, remember, and adapt to changing stimuli are a breakthrough in neuro-inspired computing. This study shows the potential of NWNs for achieving such capabilities, crucial for real-world applications like robotics and sensor edge devices~\cite{sandamirskaya2022rethinking}.

Neuromorphic processors, with their low power consumption, are set to play a crucial role in various edge-computing and edge-learning applications in autonomous systems, robotics, remote sensing, implantable, wearables, and the Internet of {\it X} Things, where the {\it X} can be medical, industrial, etc~\cite{schuman2022opportunities}.

\paragraph{\textbf{Promises of Neuromorphic AI}}

A neuromorphic device utilizing CMOS technology has been developed to detect epileptic seizures by analyzing local field potential (LFP) signals\cite{ronchini2021cmos}. This study's prospective direction allows closed-loop intervention for seizure duration reduction with SNN with a delay of 64.98$\pm$30.92 ms and only consuming \(<\)50~pW to detect ictal events. Followed by that study, NET-TEN emerged as a neuromorphic network designed to detect seizure-related brain activity using an analog approach. This technology improves upon existing neuromorphic processors by reducing the required area and power consumption, making it more feasible for use in implantable devices \cite{ronchini2023net}.

The Mayo Epilepsy Personal Assistant Device (EPAD) is a system that records objective EEG data during seizures and adjusts stimulation therapy based on seizure forecasts. However, implementing this system presents challenges such as computationally intensive forecasting algorithms, varying response timescales, and concerns regarding reliance on cloud-based solutions for certain tasks~\cite{pal2021epilepsy}.

\paragraph{\textbf{Data Security, Privacy, Dangers of these techniques Hazards and Pitfalls}}

Some neuromodulation devices detect, digitize, interpret, and act on information in neural activity systems. Systems must be developed to guard against this data being abused or hacked. Issues to be addressed include how long and where these data should be stored and who is in charge. If data can be “written to” the brain, we need systems to guard against undesirable intrusions. 

Access to data provided by a medical device can be empowering for patients. This allows them to receive reports on their health data and receive alerts for concerning events such as seizures.

Two research works focus on enhancing the security of devices related to insulin pumps for diabetic patients. In the first study, they developed an on-chip neural network system. In the second work, the researchers proposed a highly accurate and efficient deep-learning methodology to combat fake glucose dosage in vulnerable devices~\cite{rathore2017dlrt,rathore2018multi}.

Unauthorized access to the device, often called {\it brainjacking}, could allow an attacker to manipulate the stimulation parameters or even cause harm to the patient. Interference with or eavesdropping on the wireless communication between the device and external equipment could disrupt therapy or cause unintended side effects. There is also a possibility that malware or other malicious software could be introduced to the device, potentially compromising patient privacy or causing harm~\cite{pycroft2016brainjacking, pugh2018brainjacking}

Communication security in implantable medical devices (IMDs) is among the most critical issues for patient safety, and several research groups are focusing on this vital subject~\cite{surianarayanan2023convergence}. However, designing a reliable solution guaranteeing the security of IMD is conflicting with several issues related to the basis of the IMD itself and the environment surrounding the patient. These issues are dependent on, and not limited to, the device's battery life, the adaptability, availability, and the requested reliability of the secure solution.

\section{Concluding remarks}

Current challenges for designing implantable stimulation devices or electroceuticals, in general, include implant volume, safety, energy consumption, limited capacity in signal processing and the need for data telemetry~\cite{clement2019brain, yoo2021neural}. We envision that effective, responsive neuromodulation needs to be computationally self-sufficient in performing active on-chip learning to eliminate regular telemetry. Recent advancements in neuromorphic computing are critical to making our vision possible. We argue that neuromorphic computing in combination with highly low-power microelectronics for sensing~\cite{denison2007amp, qian2011micropower} and stimulation~\cite{ker2011stimulus} will enable the emergence of neuromorphic neuromodulation device as a long-term solution for intractable neurological diseases.

\section{Acknowledgement}

Luis Fernando Herbozo Contreras would like to acknowledge the partial support of the Faculty of Engineering Research Scholarship provided by the University of Sydney.  Zhaojing Huang would like to acknowledge the support of the Research Training Program (RTP) provided by the Australian Government. 
Omid Kavehei acknowledges the support provided by The University of Sydney through a SOAR Fellowship and Microsoft's support through a Microsoft AI for Accessibility grant. The author acknowledges the financial support from the Australian Research Council under Project DP230100019.

\bibliographystyle{ieeetr} 

\end{document}